\documentclass[12pt]{article}
\usepackage{amssymb,amsmath,epsfig}
\allowdisplaybreaks

\begin{document}
\title{\bf Cavity Evolution and Instability Constraints of Relativistic Interiors}

\author{Z. Yousaf \thanks{zeeshan.math@pu.edu.pk} and M. Zaeem ul Haq Bhatti \thanks{mzaeem.math@pu.edu.pk}\\
Department of Mathematics, University of the Punjab,\\
Quaid-i-Azam Campus, Lahore-54590, Pakistan.}

\date{}

\maketitle
\begin{abstract}
In this manuscript, we have identified the dynamical instability
constraints of a self-gravitating cylindrical object within the
framework of $f(R,T)$ theory of gravity. We have explored the
modified field equations and corresponding dynamical equations for
the systematic constructions of our analysis. We have imposed the
linear perturbations on metric and material variables with some
known static profile up to first order in the perturbation
parameter. The role of expansion scalar is also examined in this
scenario. The instability regimes have been discussed in the
background of Newtonian and post-Newtonian limits. We found that the
dark source terms due to the influence of modification in the
gravity model is responsible for the instability of the system.
\end{abstract}

\section{Introduction}

The global properties of our universe have been studied extensively
by the researchers within the background of general theory of
relativity (GR). Many new insights of astrophysics and cosmology
have revealed the unexpected picture of the universe. The latest
data of some reliable sources such as supernovae surveys and cosmic
microwave background radiation have put forth the dark side of the
universe. Therefore, one has to accept that current matter-energy
contents and the evolutionary picture of the universe is astonishing
and requires some explanation. During the last couple of decades,
the discussions in the standard GR due to the observational evidence
is unable to describe the key features of the present low energy
universe without imposing certain assumptions. Particularly, it
requires the inclusion of some exotic contribution in the matter
contents of the cosmos in order to study the dynamics of stars and
their clusters as well as the current accelerated expansion of the
universe.

The modified gravity approach indicates that the accelerated
expansion is due to the influence of modification in gravity for
late/early time universe. Some generalizations have been proposed
for GR since its inception, most of them have not passed the test of
time. It is worth mentioning that any reasonable gravity theory
should reduce to Newtonian (N) gravity for a slowly moving weak
source. An additional degree of freedom exist generically in any
modification of GR. There exist different modifications in the
Einstein-Hilbert (EH) action describing the dark components (i.e.,
dark energy and dark matter) of the current accelerated expanding
universe. Dark energy is used to explain the cosmic speed up and
dark matter is used to explain the emergence of large scale
structures in the universe. Thus, a wide class of gravity theories
exist to study the dark side side of the universe via the
enhancement of the gravitational force. The $f(R,T)$ gravity theory
is one of such generalization to Einstein's theory of gravity which
constitutes on the matter and geometry coupling. In this theory, the
Lagrangian for EH action includes the extra degrees of freedom along
with trace of stress energy tensor.

The exploration of instability regimes for collapse process
strengthened the study of astronomical and astrophysical theories.
The gravitational test through pulsar-timing experiments have
motivated to study the stability issue. Harada \cite{1} presented
the stability analysis in scalar tensor theory for spherically
symmetric star configuration and extracted the range of instability
from the first order derivative of coupling function. In the study
of gravitational instability theory, the amplifications in the
density perturbations are responsible for the generation of cosmic
structures in the early universe. For a time dependent mass density,
the instability for a gravitating system has been discussed in the
framework of $f(R)$ theory \cite{2}. Bamba et al. \cite{3} have
studied the matter instability describing the curvature inside the
sphere in $f(R)$ gravity theory, which is one of the most important
criteria to check the validity of any modified gravity theory. It
has been the matter of interest for relativists that dynamical
equations are developed due to the Bianchi identities and Einstein's
equations. Also, an exact evolution equation for Lagrangian can be
obtained through Bianchi identities representing the gravitational
tidal field. Sharif and Yousaf \cite{3a} have explored instability
conditions of stellar systems at both N and post-Newtonian (pN) eras
with different backgrounds in modified gravity theories.

Nojiri and Odintsov \cite{4} studied the behavior of modified
gravity on some solar systems test for Newton law corrections and
matter instabilities with the effective cosmological constant regime
in the late and early universe. Tiret and Combes \cite{5} presented
the dynamical evolution of spiral galaxies in modified dynamics,
which is compared to the  gravity with dark matter with numerical
simulations. Bogdanos and Saridakis \cite{6} explored the
perturbative instabilities by imploding scalar and tensor
perturbations within a flat background in Ho\v{r}ava gravity. Some
astrophysical test for modified gravity theories have been presented
by Jain \emph{et al.} \cite{7} using low-redshift distance
indicators to carry out tests and mainly focused on particular
stages on the evolution of supergiant and giants to observe distinct
observational signatures.

The motion of the matter can be characterized by the fluid
parameters like four-acceleration, shear tensor, expansion scalar
and vorticity tensor (which is zero for spherical stars). The
significance of shear scalar and its vanishing have been brought
forward by many researchers for self-gravitating stars. The
expansion scalar measures the change in the volume element of fluid
configuration during the evolution and its absence leads to the
formation of a cavity within the system. This is based on the reason
that during the evolution the system is expanding leading to an
increase in the volume element due to the increase in the external
boundary. The increase in the volume is compensated by the formation
of a cavity inside system by imposing the the expansion free
condition and the innermost shell would be away from the center.

Initially, Skripkin \cite{8} analyzed the appearance such kinds of
vacuum cavity during the evolution of spherically symmetric models
which has significance in the modeling of voids. Later, it was found
\cite{9} that the Skripkin model has no compatibility with the
Darmois matching conditions \cite{10}. In the same paper, they also
examine that spherical stars evolving under the expansion free
condition must have inhomogeneous energy density. The formation of
cavity using the kinematical quantities different from zero
expansion condition has also been investigated in the literature
\cite{11}. Herrera \emph{et al.} \cite{12} also explored the
instability eras for spherically symmetric collapsing model due to
zero expansion in the fluid configuration. Sharif and Bhatti
\cite{13} explored instability conditions of cylindrically symmetric
self-gravitating systems coupled with charged expansion-free
anisotropic matter distribution.

In the study of stellar structures, it is common to model the star
interior with perfect fluid which implies the same pressure in the
interior of the compact object. Some theoretical advances in recent
decades indicate the deviation from isotropic pressure particularly
in high density regimes to study their properties. Weber \cite{14}
proposed that strong magnetic fields play role for generating
pressure anisotropy inside a compact star. It is also observed that
anisotropy of pressures is present in wormholes \cite{15} or
gravastars \cite{16}, so-called exotic solutions of the field
equations. General relativistic stellar models gain crucial
significance due to the existence of pressure anisotropy in the
matter distribution \cite{17}. Sharif with his research fellows
\cite{18} investigated different effects of physical parameters on
the dynamical instability of self-gravitating collapsing stars. The
existence of various compact objects in the realm of $f(R)$ gravity
have been investigated \cite{yb1}. Recently, Yousaf et al.
\cite{ybb1} explored the importance of energy density
inhomogeneities in the study of stellar collapse.

The current paper presents a full analytical approach have been
initiated to understand the instability regimes of cylindrical
object within the physical background of $f(R,T)$ dark sources. The
format of this paper is as follows. The next section explores some
basics of $f(R,T)$ theory of gravity including the modified field
equations and kinematical quantities. In section \textbf{3}, we have
provided the perturbation scheme up to first order to analyze the
stability of our gravitating source. In section \textbf{4}, we have
found the collapse equation by exploring the dynamical equation
c.g.s unit systems. Section \textbf{5} investigates the instability
ranges under N limit with the zero expansion condition. The last
section concludes our main findings.

\section{The $f(R,T)$ Gravity and Cylindrical Systems}

The notion of $f(R,T)$ gravity as a possible modifications in the
gravitational framework of GR received much attention of
researchers. This theory provides numerous interesting results in
the field of physics and cosmology like plausible explanation to the
accelerating cosmic expansion \cite{ya3, b2a, b2b}. The main theme
of this theory is to use an algebraic general function of Ricci as
well trace of energy momentum tensor in the standard EH action. It
can be written as \cite{ya9}
\begin{equation}\label{1}
S_{f(R,T)}=\int d^4x\sqrt{-g}[f(R,T)+L_M],
\end{equation}
where $g,~T$ are the traces of metric as well as standard GR
energy-momentum tensors, respectively while $R$ is the Ricci scalar.
There exists variety $L_M$ in literature which corresponds to
particular configurations of relativistic matter distributions.
Choosing $L_M=\mu$ (where $\mu$ is the system's energy density) and
varying the above action with respect to $g_{\alpha\beta}$, the
corresponding $f(R,T)$ field equations are given as follows
\begin{equation}\label{2}
{G}_{\alpha\beta}={{T}_{\alpha\beta}}^{\textrm{eff}},
\end{equation}
where
\begin{align*}\nonumber
{{T}_{\alpha\beta}}^{\textrm{eff}}&=\left[(1+f_T(R,T))T^{(m)}_{\alpha\beta}-\mu
g_{\alpha\beta}f_T(R,T)
-\left(\frac{f(R,T)}{R}-f_R(R,T)\right)\frac{R}{2}\right.+\\\nonumber
&\left.+\left({\nabla}_\alpha{\nabla}_
\beta+g_{\alpha\beta}{\Box}\right)f_R(R,T)\right]\frac{1}{f_R(R,T)}
\end{align*}
is the effective energy-momentum tensor representing modified
version of gravitational contribution coming from $f(R,T)$ extra
degrees of freedom while ${G}_{\alpha\beta}$ is an Einstein tensor.
Further, $\nabla_\alpha$ represents covariant derivation while
$f_T(R,T),~\Box$, $f_R(R,T)$ indicate
$\frac{df(R,T)}{dT},~\nabla_\alpha\nabla^\alpha$ and
$\frac{df(R,T)}{dR}$ operators, respectively.

The system under consideration is modeled as a cylindrical stellar
object whose relativistic motion is characterized by three
dimensional timelike surface represented by $\Sigma^{(e)}$. This
boundary demarcated our manifold into two different interior and
exterior portions. These regions are denoted by $\mathcal{V}^-$ and
$\mathcal{V}^+$, respectively. The $\mathcal{V}^-$ region can be
described with the help of the following non-rotating diagonal
spacetime \cite{v33}
\begin{equation}\label{3}
ds^2_-=A^2(t,r)dt^{2}-B^2(t,r)dr^{2}-C^2(t,r)d\phi^{2}-dz^2,
\end{equation}
while spacetime for $\mathcal{V}^+$ is \cite{v34}
\begin{equation}\label{4}
ds^2_+=\left(-\frac{2M}{r}\right)d\nu^2
+2d{\nu}d{r}-r^2(d\phi^2+\zeta^2dz^2),
\end{equation}
where $M$ is a cylindrical gravitating mass, $\nu$ is the retarded
time and $\zeta$ indicates arbitrary constant. The mathematical
formula describing fluid distribution within the cylindrical
relativistic interior is \cite{v35}
\begin{equation}\label{5}
T^-_{\alpha\beta}=(\mu+P_{r})V_{\alpha}V_{\beta}-P_r
g_{\alpha\beta}+(P_{z}-P_{r})S_{\alpha}S_{\beta}+
(P_\phi-P_{r})K_{\alpha} K_{\beta},
\end{equation}
where $P_{\phi},~P_{r}$ and $P_{z}$ are stresses corresponding to
$\phi,~r$ and $z$ directions, respectively. Here $V_{\beta}$ and
$K_{\beta},~S_{\beta}$ are four velocity and four-vectors,
respectively which under the following comoving coordinate system
$$V_{\beta}=A\delta^{0}_{\beta},~S_{\beta}={\delta}^{3}_{\beta},~
K_{\beta}=C{\delta}^{2}_{\beta},$$ obey some relations. These are
given as follows
\begin{equation*}
V^{\beta}V_{\beta}=-1,\quad
K^{\beta}K_{\beta}=S^{\beta}S_{\beta}=1,\quad
V^{\beta}K_{\beta}=S^{\beta}K_{\beta}=V^{\beta}S_{\beta}=0.
\end{equation*}
The scalar variable controlling expansion and contraction of matter
distribution is known as expansion scalar. This can be obtained
through $\Theta=V^{\alpha}_{~;\alpha}$ mathematical expression. The
expansion scalar associated with cylindrically symmetric
relativsitic interior is
\begin{equation}\label{6}
\Theta=\frac{1}{A}\left(\frac{\dot{B}}{B} +\frac{\dot{C}}{C}\right),
\end{equation}
where over dot symbolizes temporal partial differenration. The
corresponding Ricci scalar is
\begin{eqnarray}\nonumber
R(t,r)&=&\frac{2}{B^2}\left[\frac{A''}{A}+\frac{C''}{C}+\frac{A'}{A}
\left(\frac{C'}{C}-\frac{B'}{B}\right)-\frac{B'C'}{BC}\right]\\\label{7}
&&-\frac{2}{A^2}\left[\frac{\ddot{B}}{B}+\frac{\ddot{C}}{C}-\frac{\dot{A}}{A}
\left(\frac{\dot{B}}{B}+\frac{\dot{C}}{C}\right)+\frac{\dot{B}\dot{C}}{BC}\right],
\end{eqnarray}
where prime means radial partial differentiation. The $f(R,T)$ field
equations (\ref{2}) for the metric (\ref{3}) give the following set
of equations
\begin{align}\label{8}
&G_{00}=\frac{A^2}{f_R}\left[\mu+\frac{R}{2}\left(\frac{f}{R}-f_R\right)+\psi_{00}\right],\quad
G_{01}=\psi_{01},\\\label{9}
&G_{11}=\frac{B^2}{f_R}\left[P_r(1+f_T)+\mu{f_T}-\frac{R}{2}\left(\frac{f}{R}-f_R\right)+\psi_{11}\right],\\\label{10}
&G_{22}=\frac{C^2}{f_R}\left[P_\phi(1+f_T)+\mu{f_T}-\frac{R}{2}\left(\frac{f}{R}-f_R\right)+\psi_{22}\right],\\\label{11}
&G_{33}=\frac{1}{f_R}\left[P_z(1+f_T)-\mu{f_T}-\frac{R}{2}\left(\frac{f}{R}-f_R\right)+\psi_{33}\right],
\end{align}
where
\begin{align}\label{12}
&\psi_{00}=\frac{\partial_{rr}f_{R}}{B^2}-\frac{\partial_tf_R}{A^2}
\left(\frac{\dot{B}}{B}+\frac{\dot{C}}{C}\right)\frac{\dot{f_{R}}}{A^2}
-\frac{\partial_rf_{R}}{B^2}\left(\frac{B'}{B}-\frac{C'}{C}\right),\\\label{13}
&\psi_{01}=\frac{1}{f_R}\left(\partial_r\partial_tf_{R}-\frac{A'}{A}\partial_tf_{R}
-\frac{\dot{B}}{B}\partial_rf_{R}\right),\\\label{14}
&\psi_{11}=\frac{\partial_t\partial_tf_R}{A^2}+\left(\frac{\dot{C}}{C}
-\frac{\dot{A}}{A}\right)\frac{\partial_tf_{R}}{A^2}
-\left(\frac{A'}{A}+\frac{C'}{C}\right)\frac{\partial_rf_{R}}{B^2},\\\label{15}
&\psi_{22}=\frac{\partial_{tt}f_{R}}{A^2}+\frac{\partial_{rr}f_{R}}{B^2}
+\left(\frac{\dot{B}}{B}-\frac{\dot{A}}{A}\right)
\frac{\partial_tf_{R}}{A^2}
+\left(\frac{B'}{B}-\frac{A'}{A}\right)\frac{\partial_rf_{R}}{B^2},\\\label{16}
&\psi_{33}=\frac{\partial_{tt}f_{R}}{A^2}-\frac{\partial_{rr}f_{R}}{B^2}+
\left(\frac{\dot{B}}{B}-\frac{\dot{A}}{A}+\frac{\dot{C}}{C}\right)\frac{\partial_tf_{R}}{A^2}
+\left(\frac{B'}{B}-\frac{A'}{A}-\frac{C'}{C}\right)\frac{\partial_rf_{R}}{B^2}.
\end{align}

Now, we are interested to formulate two equations describing the
dynamical evolution of cylindrical relativistic interiors framed
within $f(R,T)$ background. In $f(R,T)$ gravitational theory, the
divergence of stress-energy tensor is non-vanishing and is obtained
as
\begin{align}\label{17}
\nabla^{\alpha}T_{\alpha\beta}=\frac{f_T}{1-f_T}\left[
(\Theta_{\alpha\beta}+T_{\alpha\beta})\nabla^\alpha{\ln}f_T
-\frac{1}{2}g_{\alpha\beta}\nabla^\alpha{T}
+\nabla^\alpha\Theta_{\alpha\beta}\right].
\end{align}
The divergence of $f(R,T)$ energy-momentum tensor yields the
following couple of continuity equations
\begin{align}\nonumber
&\dot{\mu}\left(\frac{1+f_T+f_Rf_T}{f_R(1+f_T)}\right)-\frac{\mu}{f_R}\partial_tf_R
-\frac{B\dot{B}}{A^2f_R}(1+f_T)(\mu+P_r)-\frac{C\dot{C}}{A^2f_R}(1+f_T)(\mu+P_\phi)\\\label{18}
&+\frac{2\mu}{1+f_T}\partial_tf_T+\frac{\partial_tT}{2(1+f_T)}+D_0=0,
\\\nonumber
&\frac{P'_{r}}{f_R}+\frac{P_{r}}{f_R}\left\{\partial_rf_T-\frac{(1+f_T)\partial_rf_R}{f_R}\right\}
-\frac{AA'}{A^2f_R}(1+f_T)(\mu+P_r)+\frac{\mu'}{f_R}+\frac{\mu}{f_R}\\\nonumber
&\times\left\{\partial_rf_T-\frac{f_T\partial_rf_R}{f_R}\right\}-\frac{(\mu-P_r)}{(1+f_T)}
\partial_rf_T+\frac{f_T}{(1+f_T)}\left(\mu'+\frac{T'}{2}\right)+\frac{C'}{CB^2f_R}\\\label{19}
&(1+f_T)(P_r-P_\phi)+D_1=0,
\end{align}
These are the required couple of dynamical equations obtained
through contracted Bianchi identities of the $f(R,T)$ effective
stress energy tensor. It is well-known that these dynamical
equations assist enough to help to analyze the dynamical evolution
of stellar system collapse with the passage of time. This also help
to explore total energy variation within the collapsing celestial
self-gravitating systems in regard with time and adjacent
boundaries. MacCallum et al. \cite{mac1} gave a nice way to discuss
perturbed boundary conditions in joining matter filled interior with
the asymptotically flat vacuum exterior solution. In the above
equations, the quantities $D_0$ and $D_1$ are functions of $t$ and
$r$ and representing extra curvature dark source terms emerging from
$f(R,T)$ gravitational field. The quantities $D_0$ and $D_1$
describe $f(R)$ corrections in the energy variations of the
collapsing cylindrical relativistic interior associated with time
and adjacent surfaces, respectively. These corrections are given in
Appendix \textbf{A}.

The matter quantity of cylindrical collapsing stellar geometry can
be defined through gravitational C-energy which was proposed by
Thorne \cite{v35a}. This is obtained as
\begin{equation}\label{20}
m(t,r)=\left\{1-\left(\frac{C'}{B}\right)^2-\left(\frac{\dot{C}}{A}
\right)^2\right\}\frac{l}{8},
\end{equation}
where $l$ indicates specific cylindrical length. Before calculating
its variations among adjacent surfaces of cylindrical anisotropic
fluid distribution, we shall define some operators. The proper and
radial derivative operators are defined as follows
\begin{eqnarray}\label{21}
D_{T}=\frac{1}{A} \frac{\partial}{\partial t},\quad
D_{C}=\frac{1}{C'}\frac{\partial}{\partial r}.
\end{eqnarray}
The relativistic velocity of the collapsing stellar interior can be
obtained with the help of proper derivative operator as
\begin{eqnarray}\label{22}
U=D_{T}C=\frac{\dot{C}}{A}.
\end{eqnarray}
Using Eqs.(\ref{20}) and (\ref{22}), we obtain
\begin{eqnarray}\label{23}
\tilde{\mathrm{E}}\equiv\frac{C'}{B}=\left[1-\frac{8}{l}m(t,r)+U^{2}
\right]^{1/2}.
\end{eqnarray}
Next, from Eqs.(\ref{20}), (\ref{21}) and (\ref{23}), it follows
that
\begin{align}\label{24}
D_{C}m&=\frac{l}{4f_{R}}\left[\mu+
\frac{R}{2}\left(\frac{f}{R}-f_R\right)+\psi_{00}-
\frac{\psi_{01}}{BA}\frac{U}{\tilde{\mathrm{E}}}\right]C.
\end{align}
This equation tells us provides the total energy variation between
adjacent surfaces within the matter configuration. From the above
equation it is evident that the first part of the right hand side is
due to energy density and $f(R)$ higher curvature quantities. The
last term in the above equation is
$(\overset{~~~(H)}{T_{01}}/BA)(U/\tilde{\mathrm{E}})<0$. In this
term, the quantity $U$ represents collapsing matter velocity. It is
well-known that $U$ is less than zero for the collapsing fluid
models. Thus, the last term (containing $U$ and non-attractive
$f(R)$ corrections) lessens fluid energy influences in the
evolutionary system phases. Equation (\ref{16}) upon integration
yields
\begin{align}\label{25}
m=\frac{1}{4}\int^C_{0}\left[\frac{lC}{f_{R}}\left(\mu+
\frac{R}{2}\left(\frac{f}{R}-f_R\right)+\psi_{00}-
\frac{\psi_{01}}{BA}\frac{U}{\tilde{\mathrm{E}}}\right)\right]dC.
\end{align}

It is known that zero expansion leads to the emergence of boundary
surfaces in which one (outer) demarcates the relativistic interior
fluid from the exterior spacetime while the second (inner) separates
Minkowskian core from the matter distribution. Under null expansion
scalar framework, the relativistic fluid evolves without being
compressed. For example, during expansion self-gravitating system,
the increase in volume of matter configuration leads to expansion of
the external boundary surface which can be counterbalanced by a
similar expansion of the internal surface to make $\Theta$ zero.
Thus, zero expansion scalar triggers evolution of relativistic
system in such a way that the inner most shell moves away from the
central point thereby causing the emergence of vacuum core. Due to
this zero expansion, matter sources could be effective for the voids
explanation. Voids are, roughly speaking, underdense areas
incorporating substantial amount of information on the cosmological
environment \cite{v1}. Voids provide a reliable guide to study the
large scale cosmic structure formation. They are more rich in
modified gravity \cite{v2} as compared to GR as this extended
gravity theory is more likely to host large structures with smaller
radii.

The continuity of Eqs.(\ref{3}) and (\ref{4}) over $\Sigma^{(e)}$
can be obtained by using Darmois matching conditions \cite{v36}
\begin{align}\label{26}
&m(t,r)-M\overset{\Sigma^{(e)}}=\frac{l}{8}, \quad
l\overset{\Sigma^{(e)}} =4C,\\\label{27} &
P_r\overset{\Sigma^{(e)}}=
\left[\mu{f_T}-\frac{f-Rf_R}{2}\right]\left[1+\frac{(1+f_T)}{f_R}\right]^{-1},
\end{align}
where $\overset{\Sigma^{(e)}}=$ means that measurements are
performed over outer hypersurface. The matching conditions over
${\Sigma^{(i)}}$ lead to
\begin{equation}\label{28}
m(t,r)\overset{\Sigma^{(i)}}{=}0,\quad
P_r\overset{\Sigma^{(i)}}=\left(\mu{f_T}-\frac{f-Rf_R}{2}\right)\left(1+\frac{(1+f_T)}{f_R}\right)^{-1}.
\end{equation}

To present $f(R,T)$ gravity as a cosmologically and theoretically
consistent theory, the selection of its models is very important.
Here, we consider particular class of models as follows
\begin{equation}\label{29n}
f(R,T)=f_1(R)+f_2(R)f_3(T).
\end{equation}
This model involves the explicit non-minimal curvature matter
coupling. We now consider $f(R,T)$ power law type model given by
\begin{equation}\label{29}
f(R)=R+\lambda R^2T^2.
\end{equation}
Such functional $f(R,T)$ configurations match to the Lagrangian form
mentioned in Eq.(\ref{29n}). Here, we take $f_1(R)=R,~f_2(R)=R^2$
and $f_3(T)=T^2$. All GR solutions can be obtained by taking limit
$\lambda\rightarrow0$.

\section{Perturbation Scheme}

Perturbation theory gives us a mathematical technique that assists
enough to find an approximate solution of a differential equation.
After applying perturbation scheme, one can break corresponding
equations into ``solvable/static" and ``perturbed" parts. The impact
of perturbed terms in the equation keeps on decreasing, controlled
by the perturbation parameter. Here, we consider perturbation scheme
that was proposed by Herrera \textit{et al.} \cite{her1a}. In this
scheme, we take $\alpha$ to be perturbation parameter with
$\alpha\in[0,1]$ and consider effects upto first order. We assume
that the system initially is in the state of hydrostatic
equilibrium. Due to this cylindrical scale factors as well as fluid
variables are independent of temporal coordinate. We shall represent
static configurations of corresponding variables by zero subscript.
Upon perturbations, all these structural variables depend upon the
same time dependence $\eta(t)$, which eventually gives same time
dependence to Ricci scalar. The perturbation scheme is
\begin{eqnarray}\label{30}
A(t,r)&=&A_0(r)+{\alpha}\eta(t)a(r),\\\label{31}
B(t,r)&=&B_0(r)+{\alpha}\eta(t)b(r),\\\label{32}
C(t,r)&=&C_0(r)+{\alpha}\eta(t){c}(r),\\\label{33}
R(t,r)&=&R_0(r)+\alpha T(t)e(r),\\\label{34}
P_{\phi}(t,r)&=&P_{\phi0}(r)+{\alpha}{\bar{P_{\phi}}}(t,r),\\\label{35}
P_{r}(t,r)&=&P_{{r}0}(r)+{\alpha}{\bar{P_{r}}}(t,r),\\\label{36}
\mu(t,r)&=&\mu_0(r)+\alpha{\bar{\mu}}(t,r), \\\label{37}
P_{\bot}(t,r)&=&P_{\bot0}(r)+{\alpha}{\bar{P_{\bot}}}(t,r),\\\label{38}
m(t,r)&=&m_0(r)+\alpha{\bar{m}}(t,r),\\\label{39}
f(t,r)&=&R_0(1+{\lambda}T_0^2R_0)+\alpha
\eta(t)e(r)(1+2{\lambda}R_0T_0^2),\\\label{40}
f_R(t,r)&=&(1+2{\lambda}R_0T_0^2)+2{\alpha}\eta(t){\lambda}
e(r)T_0^2,\\\label{41} \Theta(t,r)&=&\alpha{\bar{\Theta}}(t,r),
\end{eqnarray}
where $R_0$ is the static form of Ricci invariant whose value is
\begin{equation*}
R_0(r)=\frac{2}{B_0^2}\left[\frac{A''_0}{A_0}+\frac{A'_0}{A_0}
\left(\frac{1}{r}-\frac{B'_0}{B_0}\right)-\frac{B'_0}{B_0r}\right],
\end{equation*}
while its perturbed form is
\begin{align*}
&-\eta
e=\frac{2\ddot{\eta}}{A_0^2}\left(\frac{b}{B_0}+\frac{c}{r}\right)
+\eta\left[\frac{4b}{B_0^3}R_0
-\frac{2}{B_0^2}\left\{\frac{a''}{A_0}-\frac{aA''_0}{A_0^2}
+\frac{c''}{r}-\frac{A'_0}{A_0}\left\{\left(\frac{b}{B_0}\right)'
\right.\right.\right.\\\nonumber
&-\left.\left.\left.\left(\frac{c}{r}\right)'\right\}+\left(\frac{a}{A_0}\right)'
\left(\frac{1}{r}-\frac{B'_0}{B_0}\right)-\frac{B'_0}{B_0}
\left(\frac{c}{r}\right)'-\frac{1}{r}\left(\frac{b}{B_0}\right)'
\right\}\right].
\end{align*}
The $f(R,T)$ field equations (\ref{8})-(\ref{11}) under static
background with $C_0=r$ turn out to be
\begin{align}\label{42}
&\frac{1}{rB_0^2}\times\frac{B'_0}{B_0}=\frac{1}{(1+2{\lambda}R_0T_0^2)}\left[\mu_0-\frac{\lambda}{2}
R_0^2T_0^2+\overset{(S)}{\psi_{00}}\right],
\\\label{43}
&\frac{1}{rB^2_0}\frac{A'_0}{A_0}=\frac{1}{(1+2{\lambda}R_0T_0^2)}\left[(P_{r0}
+\mu_0)(1+2{\lambda}R_0^2T_0)+\frac{\lambda}{2}R_0^2T_0^2+\overset{(S)}{\psi_{11}}
\right],\\\nonumber
&\frac{A'_0}{B_0A_0}\left(\frac{B'_0}{B^2_0}+\frac{A''_0}{B_0A'_0}\right)
=\frac{1}{(1+2{\lambda}R_0T_0^2)}\left[(P_{\phi0}+\mu_0)(1+2{\lambda}R_0^2T_0)\right.\\\label{44}
&
+\left.\frac{\lambda}{2}R_0^2T_0^2+\overset{(S)}{\psi_{22}}\right],\\\nonumber
&\left(\frac{A'_0}{A_0r}+\frac{A''_0}{A_0}\right)\frac{1}{B_0^2}-\frac{B'_0}{B_0^3}
\left(\frac{1}{r}+\frac{A'_0}{A_0}\right) =\frac{1}{(1+2{\lambda}
R_0T_0^2)}\left[(P_{z0}-\mu_0)(1+2{\lambda}\right.\\\label{45}
&\left.\times R_0^2T_0)
+\frac{\lambda}{2}R_0^2T_0^2+\overset{(S)}{\psi_{33}}\right],
\end{align}
where $\overset{(S)}{\psi_{ii}}$ indicate static configurations of
the corresponding dark source components and are given by
\begin{align}\nonumber
\overset{(S)}{\psi_{00}}&=\frac{2\lambda
T_0}{B_0^2}\left[2R_0T'_0+T_0R''_0+2R_0T''_0+2R_0\frac{T'^2_0}{T_0}
+\left(\frac{1}{r}-\frac{B'_0}{B_0}\right)(T_0R'_0\right.\\\nonumber
&+\left.2R_0T_0')\right],\\\nonumber
\overset{(S)}{\psi_{11}}&=-\frac{2\lambda{T_0}}{B_0^2}\left(\frac{A'_0}{A_0}+\frac{1}{r}\right)(T_0R'_0
+2R_0T'_0),\\\nonumber
\overset{(S)}{\psi_{22}}&=\frac{2\lambda{T_0}}{B_0^2}\left[
2R_0T'_0+T_0R''_0+2R_0T''_0+2R_0\frac{T'^2_0}{T_0}
+\left(\frac{B'_0}{B_0}-\frac{A_0'}{A_0}\right)(T_0R'_0\right.\\\nonumber
&+\left.2R_0T'_0)\right],\\\nonumber
\overset{(S)}{\psi_{33}}&=\frac{2\lambda
T_0}{B_0^2}\left[(T_0R'_0+2R_0T'_0)\left(\frac{B'_0}{B_0}-\frac{A'_0}{A_0}-\frac{1}{r}\right)-
2R_0T'_0-T_0R''_0-2R_0T''_0\right.\\\nonumber
&-\left.2R_0\frac{T'^2_0}{T_0}\right].
\end{align}
Equations (\ref{8})-(\ref{10}) under non-hydrostatic state take the
following form
\begin{align}\label{46}
&\bar{\mu}=\chi_2\eta,\\\nonumber &\bar{\mu}+\bar{P_r}=\eta\chi_3
-\frac{\overset{(P_1)}{\psi_{00}}\dot{\eta}}{(1+2{\lambda}R_0^2T_0)}
-\frac{\ddot{\eta}}{(1+2{\lambda}R_0^2T_0)}\left\{\overset{(P_2)}{\psi_{00}}
+\frac{c}{rA_0^2}(1+2{\lambda}R_0T_0^2)\right\}\\\label{47}
&\times\left(\frac{2b}{B_0^2}+2e\lambda T_0^2\right),\\\label{48}
&\bar{\mu}+\bar{P_\phi}=\eta\chi_5-\frac{\ddot{\eta}}{(1+2{\lambda}R_0^2T_0)}
\left[\frac{b}{A_0^2B_0}(1+2{\lambda}R_0T_0^2)+\overset{(P_2)}{\psi_{22}}\right],
\end{align}
where $\chi_i's$ are dark source terms coming due to $f(R,T)$
gravity. These terms contain static combinations of metric variables
and are mentioned in Appendix \textbf{A}. After applying our radial
perturbation approach, we have seen that first dynamical equation
(\ref{18}) is trivially obeyed while the second dynamical equation
(\ref{19}) turns out to be
\begin{align}\nonumber
&\frac{P'_{r0}}{(1+2\lambda R_0T_0^2)}+\frac{2\lambda
P_{r0}}{(1+2\lambda
R_0T_0^2)}\left[R_0(2T_0R'_0+R_0T'_0)-\frac{T_0(T_0R'_0+2R_0T'_0)}{(1+2\lambda
R_0T_0^2)}\right.\\\nonumber &\left.\times(1+2\lambda
R_0^2T_0)\right]\left[\frac{b}{B_0}+\frac{a}{A_0}+\frac{2e\lambda
T_0^2}{(1+2{\lambda}R_0T_0^2)}\right]+\frac{\mu'_0}{(1+2\lambda
R_0T_0^2)}+\frac{2\lambda\mu_0}{(1+2\lambda R_0T_0^2)}\\\nonumber
&\times\left[R_0(2T_0R'_0+R_0T'_0)-\frac{2\lambda
T_0^2R_0^2(T_0R'_0+2R_0T'_0)}{(1+2\lambda R_0T_0^2)}\right]-2\lambda
R_0\frac{(2T_0R'_0+2R_0T'_0)}{(1+2\lambda R_0^2T_0)}\\\nonumber
&\times(\mu_0-P_{r0})+\frac{2\lambda R_0^2T_0}{(1+2\lambda
R_0T_0^2)}\left(\mu_0'+\frac{T'_0}{2}\right)+\frac{r(1+2\lambda
R_0^2T_0)}{B_0^2(1+2\lambda R_0T_0^2)}(P_{r0}-P_{\phi0})\\\label{49}
&-\frac{A_0A'_0}{B_0^2(1+2\lambda
R_0T_0^2)}(\mu_0+P_{r0})(1+2\lambda R_0^2T_0)+D_{1S}=0,
\end{align}
where $D_{1S}$ is the static form of $D_1$ and is found to be
\begin{align}\nonumber
D_{1S}&=\overset{(S)}{\psi_{11,1}}-\frac{A_0A'_0}{B_0^2(1+2\lambda
R_0T_0^2)}\left(\overset{(S)}{\psi_{00}}+\overset{(S)}{\psi_{11}}\right)+
\frac{\lambda}{2}(R_0^2T_0^2)'
+\frac{r\left(\overset{(S)}{\psi_{11}}+\overset{(S)}{\psi_{22}}\right)}{B_0^2(1+2\lambda
R_0T_0^2)}.
\end{align}
The static as well as non-static portions of cylindrical
relativistic C-energy function turn out to be
\begin{align}\label{50}
&m_0=\frac{l}{8}\left(1-\frac{1}{B_0^2}\right),\quad
\bar{m}=\frac{l}{4}\left(\frac{b}{B_0}-c'\right)\frac{\eta}{B_0^2},
\end{align}
the expansion scalar in its perturbed formulation is
\begin{equation}\label{51}
\bar{\Theta}=\left(\frac{b}{B_0}+\frac{c}{r}\right)\frac{\dot{\eta}}{A_0}.
\end{equation}
The equation relating $b$ and $B_0$ is found after perturbing
Eq.(\ref{8}) as
\begin{eqnarray}\nonumber
\frac{b}{B_0}=r\left(\psi_{01}
-\frac{cA_0'}{rA_0}+\frac{c'}{r}\right).
\end{eqnarray}
The first and second conservation laws (\ref{18}) and (\ref{19})
after using perturbation scheme provide the following non-static
perturbed distributions as
\begin{align}\label{52}
&\bar{\dot{\mu}}+\chi_1(r)\dot{\eta}=0,
\\\nonumber
&\frac{\bar{P'_r}}{(1+2\lambda
R_0^2T_0)}+\frac{2\lambda\bar{P_r}}{(1+2\lambda
R_0^2T_0)}\left[R_0(2T_0R'_0+R_0T'_0)-T_0(T_0R'_0+2R_0T'_0)\right.\\\nonumber
&\left.\frac{(1+2\lambda R_0^2T_0)}{(1+2\lambda
R_0T_0^2)}\right]-(\bar{\mu}+\bar{P_r})
\frac{A_0A'_0}{B^2_0}+\frac{\bar{\mu'}}{(1+2\lambda R_0^2T_0)}
+\frac{2\lambda\bar{\mu}}{(1+2\lambda
R_0^2T_0)}\left[R_0(2T_0\right.\\\nonumber
&\times\left.R'_0+R_0T'_0)-\frac{2\lambda R_0^2T_0^2}{(1+2\lambda
R_0T_0^2)}\right]+\frac{2\lambda R_0^2T_0\bar{\mu'}}{(1+2\lambda
R_0^2T_0)}-2\lambda R_0\frac{(2T_0R'_0+R_0T'_0)}{(1+2\lambda
R_0^2T_0)}\\\nonumber
&\times(\bar{\mu}-\bar{P_r})+\frac{r}{B_0^2}(\bar{P_r}-\bar{P_\phi})\frac{(1+2\lambda
R_0^2T_0)}{(1+2\lambda R_0T_0^2)} -\frac{A_0A'_0}{B_0^2(1+2\lambda
R_0T_0^2)}
\left(\dot{\eta}\overset{(P_1)}{\psi_{11}}+\ddot{\eta}\overset{(P_2)}{\psi_{11}}\right)\\\label{53}
&+\frac{r}{B_0^2(1+2\lambda
R_0T_0^2)}\left\{\ddot{\eta}\left(\overset{(P_2)}{\psi_{11}}-\overset{(P_2)}{\psi_{22}}\right)
+\dot{\eta}\overset{(P_1)}{\psi_{11}}\right\}+\frac{2e\lambda
T_0\eta P'_{r0}}{(1+2\lambda R_0T_0^2)^2}+\eta D_3=0,
\end{align}
where $\chi_1$ constitutes gravitational effects coming from
effective energy density of the cylindrical stellar objects, while
$D_3$ contains $f(R,T)$ higher curvature corrections. These terms
are mentioned in Appendix \textbf{A}. Equations (\ref{52}) and
(\ref{53}), known as dynamical equations, would be very useful in
the discussion of collapsing behavior of relativistic stellar
interiors.

The matching condition (\ref{27}) in account of perturbation
provides
\begin{align}\label{54}
P_{r0}&\overset{\Sigma^{(e)}}=-\left(1+\frac{(1+2\lambda
R_0^2T_0)}{(1+2\lambda R_0T_0^2)}\right)^{-1}\left[\mu_0(1+2\lambda
R_0^2T_0)+\frac{\lambda}{2} R_0^2T_0^2\right], \\\nonumber
\bar{P}_{r}&\overset{\Sigma^{(e)}}=\left(1+\frac{(1+2\lambda
R_0^2T_0)}{(1+2\lambda R_0T_0^2)}\right)^{-1}[(2e\lambda
R_0^2\mu_0+e\lambda R_0T_0^2)\eta+\bar{\mu}(1+2\lambda
R_0^2T_0)]\\\label{55} &+\frac{2e\lambda P_{r0}\eta}{(1+2\lambda
R_0T_0^2)} \left[R_0^2+\frac{T_0^2(1+2\lambda R_0^2T_0)}{(1+2\lambda
R_0T_0^2)}\right]\left[1+\frac{(1+2\lambda R_0^2T_0)}{(1+2\lambda
R_0T_0^2)}\right].
\end{align}
Using (00) field equation in the second of the above equation, we
get
\begin{align}\label{56}
\bar{P}_{r}&\overset{\Sigma^{(e)}}=\chi_4\eta.
\end{align}
where
\begin{align}\nonumber
\chi_4&\overset{\Sigma^{(e)}}=e\lambda R_0\left(1+\frac{(1+2\lambda
R_0^2T_0)}{(1+2\lambda
R_0T_0^2)}\right)^{-1}(2\mu_0R_0+T_0^2)+\frac{2e\lambda
P_{r0}}{(1+2\lambda R_0T_0^2)}\\\nonumber &
\times\left[R_0^2+\frac{T_0^2(1+2\lambda R_0^2T_0)}{(1+2\lambda
R_0T_0^2)}\right]\left[1+\frac{(1+2\lambda R_0^2T_0)}{(1+2\lambda
R_0T_0^2)}\right]+\chi_2(1+2\lambda R_0^2T_0)\\\nonumber &\times
\left(1+\frac{(1+2\lambda R_0^2T_0)}{(1+2\lambda
R_0T_0^2)}\right)^{-1}.
\end{align}
Equations (\ref{47}) and (\ref{56}), after some manipulation, gives
the following second order partial differential equation
\begin{equation}\label{57}
\gamma_1\ddot{\eta}+\gamma_2\dot{\eta}+\gamma_3\eta\overset{\Sigma^{(e)}}=0,
\end{equation}
where
\begin{align}\nonumber
\gamma_1&=\frac{1}{(1+2\lambda
R_0^2T_0)}\left\{\overset{(P_2)}{\psi_{11}}+\frac{c(1+2\lambda
R_0T_0^2)}{rA_0^2}\right\},\quad
\gamma_2=\frac{\overset{(P_1)}{\psi_{11}}}{(1+2\lambda
R_0^2T_0)},\\\nonumber \gamma_3&=\chi_4-\chi_3+\chi_2.
\end{align}
Equation (\ref{57}) has two solutions with two different behaviors
and these behaviors are totally independent of each other. In this
paper, we are interested to find the unstable constraints of
evolving cylindrical compact object in modified gravity. Further, it
is mentioned earlier that our system was initially in complete
hydrostatic equilibrium. Then, it enters into the collapsing phase
by reducing its areal radius. Therefore, we now restrict our
perturbations in such a way that all radial perturbed functions,
i.e., $a,~b,~c$ and $e$, are positive definite, which eventually
making $\omega_{\Sigma^{(e)}}^2>0$. The solution associated with
Eq.(\ref{57}) is obtained as follows
\begin{equation}\label{58}
\eta=-\exp{(\omega_{\Sigma^{(e)}}t)}, \quad \textmd{where} \quad
\omega_{\Sigma^{(e)}}=\frac{-\gamma_2+\sqrt{\gamma_2^2-4\gamma_1\gamma_3}}{2\gamma_1}.
\end{equation}

\section{N \& pN Terms and Collapse Equation}

In this section, we shall express second dynamical equation into
centimeter-gram-second (cgs) units and then indicate terms relating
to N, pN and parameterized post Newtonian (ppN) epochs. This would
be done by expanding cgs second dynamical equation upto
$O(\frac{1}{\mathcal{C}^4})$, where $\mathcal{C}$ indicates light
speed. For N and pN epochs, we shall consider the following
approximations
\begin{equation}\label{59}
\mu_0\gg P_{j0},\quad A_0=1-\frac{m_0\mathcal{G}}{r\mathcal{C}^2},
\quad B_0=1+\frac{m_0\mathcal{G}}{r\mathcal{C}^2},
\end{equation}
where $\mathcal{G}$ is the gravitational constant. Equation
(\ref{44}) gives us following peculiar form of double derivative of
$A_0$ with respect to radial coordinate
\begin{equation}\label{60}
\frac{A''_0}{A_0}=-\frac{A'_0B'_0}{A_0B_0}+\frac{B_0^2}{(1+2\lambda
R_0T_0^2)}\left[(\mu_0+P_{\phi0})(1+2\lambda R_0^2T_0)
+\frac{\lambda}{2}R_0^2T_0^2+\overset{(S)}{\psi_{22}}\right].
\end{equation}
Equations (\ref{43}) and (\ref{50}) provide the following first
radial derivatives of $A_0$ and $B_0$ as
\begin{align}\label{61}
\frac{B'_0}{B_0}&=\frac{4m'_0}{(l-8m_0)},\\\label{62}
\frac{A'_0}{A_0}&=\frac{2r^2l(\mu_0+P_{r0})(1+2\lambda
R_0^2T_0)+\lambda R_0^2T_0^2r^2l-4(l-8m_0)\lambda T_0^2}
{2r(l-8m_0)(1+2\lambda R_0T_0^2+2\lambda
rT_0)}=\varphi\textmd{(say)}.
\end{align}
We use Eqs.(\ref{59}), (\ref{60}) and (\ref{62}) in static form of
second dynamical equation (\ref{49}) which after converting into cgs
system turns out to be
\begin{align}\nonumber
&P'_{r0}=(1+2\lambda
\mathcal{C}^{-4}\mathcal{G}R_0^2T_0)(P_{\phi0}-P_{r0})\frac{1}{r}\frac{\mathcal{G}}{\mathcal{C}^2}
\left(\frac{r}{\mathcal{G}\mathcal{C}^2}-2m_0\right)-2\lambda
P_{r0}\left[\frac{R_0\mathcal{G}}{\mathcal{C}^4}(2T_0R'_0\right.\\\nonumber
&\left.+R_0T'_0)-\frac{\mathcal{C}^4}{\mathcal{G}}T_0(T_0R'_0+2R_0T'_0)\frac{(1+2\lambda
\mathcal{C}^{-4}\mathcal{G}R_0^2T_0)}{(1+2\lambda
\mathcal{C}^4\mathcal{G}^{-1}R_0T_0^2)}\right]-\mu'_0\mathcal{C}^2-2\lambda\mu_0\mathcal{G}\mathcal{C}^{-2}
\left[R_0\right.\\\nonumber
&\times\left.(2T_0R'_0+R_0T'_0)+\frac{2\lambda
R_0^2T_0^2(R'_0T_0+2R_0T'_0)}{(1+2\lambda
\mathcal{C}^{4}\mathcal{G}^{-1}R_0T_0^2)}+2\lambda
R_0\left(\frac{\mathcal{G}\mu_0}{\mathcal{C}^2}-\frac{P_{r0}\mathcal{G}}{\mathcal{C}^4}\right)
(2T_0R'_0\right.\\\nonumber &\left.+R_0T'_0)-2\lambda
R_0^2T_0\left(\frac{\mathcal{G}\mu'_0}{\mathcal{C}^2}+\frac{\mathcal{C}^4T'_0}{2\mathcal{G}}\right)\right]
+\varphi|_\mathcal{G}\left[\left(\frac{r^2-4rm_0\mathcal{G}\mathcal{C}^{-2}}{r^2}\right)\left\{
\frac{2\lambda\mathcal{C}^8T_0}{\mathcal{G}^2r}\left(r\right.\right.\right.\\\nonumber
&\left.\left.\left.-\frac{\mathcal{G}m_0}{\mathcal{C}^2}
\right)\left(2R_0T'_0+T_0R''_0+2R_0T''_0+2R_0\frac{T'^2_0}{T_0}\right)
+\mathcal{C}^2\left(\mu_0+\frac{P_{r0}}{\mathcal{C}^2}\right)
(1+2\lambda \mathcal{C}^{-4}\right.\right.\\\nonumber
&\times\left.\left.\mathcal{G}R_0^2T_0)+\frac{2\lambda
\mathcal{C}^8T_0}{r\mathcal{G}^2}\left(r-\frac{2\mathcal{G}m_0}{\mathcal{C}^2}\right)
\left(\frac{4\mathcal{G}m_0\mathcal{C}^{-2}}{l-8\mathcal{G}m_0\mathcal{C}^{-2}}
-\varphi|_\mathcal{G}\right)(T_0R'_0+2R_0T'_0)\right\}\right.\\\nonumber
&\left.-\frac{4\lambda\mathcal{C}^8T_0}{\mathcal{G}^2}
\left(r-\frac{2\mathcal{G}m_0}{\mathcal{C}^2}\right)(T_0R'_0+2R_0T'_0)\right]
+\frac{2\lambda\mathcal{C}^4}{\mathcal{G}}(1+2\lambda
\mathcal{C}^{4}\mathcal{G}^{-1}R_0T_0^2)\left[\left(1\right.\right.\\\nonumber
&-\left.\left.
\frac{2\mathcal{G}m_0}{r\mathcal{C}^{2}}\right)(T_0R'_0+2R_0T'_0)
\left(\frac{1}{r}+\varphi|_\mathcal{G}\right)\right]_{,1}
+2(r-4\mathcal{G}m_0\mathcal{C}^{-2})\lambda\mathcal{C}^8
\mathcal{G}^{-2}T_0\left[\frac{1}{r}\right.\\\label{g1}
&\left.+\frac{4\mathcal{G}\mathcal{C}^{-2}m'_0}{(l-8\mathcal{G}m_0\mathcal{C}^{-2})}+4R_0T'_0
+T_0R''_0+2R_0T''_0+\frac{2R_0T'^2_0}{T_0}\right].
\end{align}
It is worthy to mention that expansion of the above equation provide
terms relating to some specific eras with details as follows
\begin{align*}
&\textmd{terms of}~O(\mathcal{C}^0)~\textmd{indicates contribution
at N era},\\\nonumber &\textmd{terms
of}~O\left(\frac{1}{\mathcal{C}^2}\right)~\textmd{indicates
contribution at N era} ,\\\nonumber &\textmd{terms
of}~O\left(\frac{1}{\mathcal{C}^4}\right)~\textmd{indicates
contribution at N era}.
\end{align*}
The complete expansion of above equation upto
$O({\mathcal{C}^{-4}})$ is described in Appendix \textbf{A}.

The $f(R,T)$ field equations are filled with complicated derivatives
of radial and temporal coordinates, the exploration of their generic
solutions is a painstaking task that yet now has not been
accomplished. However, certain restrictions with some physical
background could assists enough to find their solution. In this
perspective, we consider expansion-free evolution of cylindrical
compact objects against linear perturbation. The expansion-free
condition can be obtained from Eq.(\ref{51}) as follows
\begin{equation}\label{63}
\frac{b}{B_0}=-\frac{c}{r}.
\end{equation}
We are now interested to calculate $f(R,T)$ expansion-free
cylindrical collapse equation. This would be furnished by
considering the second perturbed dynamical equation along with the
supposition that cylindrical evolution is supported by null
expansion scalar background. Thus, using Eq.(\ref{53}), junction
conditions (\ref{54}), (\ref{55}), (\ref{58}) and above
expansion-free constraint, we obtain the following modified collapse
equation over the exterior boundary surface
\begin{align}\nonumber
&\frac{\chi'_4\eta}{(1+2\lambda
R_0^2T_0)}+\frac{2\lambda\chi_4\eta}{(1+2\lambda
R_0^2T_0)}\left[R_0(2T_0R'_0+R_0T'_0)-T_0(T_0R'_0+2R_0T'_0)\right.\\\nonumber
&\times\left.\frac{(1+2\lambda R_0^2T_0)}{(1+2\lambda
R_0T_0^2)}\right]-(\chi_2+\chi_4)\eta
\frac{A_0A'_0}{B^2_0}+\frac{\chi'_2\eta}{(1+2\lambda R_0^2T_0)}
+\frac{2\lambda\chi_2}{(1+2\lambda
R_0^2T_0)}\left[R_0\right.\\\nonumber
&\times\left.(2T_0R'_0+R_0T'_0)-\frac{2\lambda
R_0^2T_0^2}{(1+2\lambda R_0T_0^2)}\right]\eta+\frac{2\lambda
R_0^2T_0\chi'_2\eta}{(1+2\lambda
R_0^2T_0)}-\frac{(2T_0R'_0+R_0T'_0)}{(1+2\lambda
R_0^2T_0)}\\\nonumber &\times2\lambda
R_0(\chi_2-\chi_4)\eta+\frac{r}{B_0^2}
\left\{(\chi_4-\chi_5-\chi_2)-\frac{\omega^2}{(1+2\lambda
R_0^2T_0)}\left(\frac{b}{B_0^2r^2}(1+2\lambda\right.\right.\\\label{64}
&\times\left.\left.
R_0T_0^2)+\overset{(P_2)}{\psi_{22}}\right)\right\}
\frac{(1+2\lambda R_0^2T_0)}{(1+2\lambda
R_0T_0^2)}\eta=(\Phi+\Omega_{e})\eta,
\end{align}
where quantities $\Phi$ and $\Omega_{e}$ are described in Appendix
\textbf{A}.  It is worthy to stress that in the above equation, the
term $\Phi$ describes the gravitational contribution in
expansion-free systems. This contribution is same as in expansion
cylindrical systems evolution. However, the quantity $\Omega_e$ is
responsible for the emergence of cylindrical central Minkowskian
core during system evolution. The gravitational effects coming from
$\Omega_e$ causes the naked singularity appearance during
cylindrical collapse. This comes from the fact that the quantity
$\Omega_e$ contains all those terms that have been evaluated by
performing expansion-free condition. When the system fluid moves
inward, during collapsing phenomenon, with null expansion rate,
there will be a blowup of the shearing scalar at the center. It is
well-known from the work of Joshi \emph{et al.} \cite{j11} is that
strong shear could produce hindrances in the appearance of apparent
horizon, thus producing a platform for naked singularity formation.
Thus, $\Omega_e$ gave a way to discuss naked singularity emergence
in a simple way. On making $\frac{b}{B_0}\neq-\frac{c}{r}$, one can
remove all of the expansion-free effects in the above collapse
equation.

\section{Instability Constraints at Both N \& pN Epochs}

In this section, we explore unstable regions of collapsing
cylindrical stellar model at both N and pN eras. Under pN limits,
Eq.(\ref{64}) in relativistic units yields
\begin{align}\nonumber
&\eta\left[\frac{\chi'_4}{(1+2\lambda
R_0T_0^2)}+\frac{2\lambda\chi_4}{(1+2\lambda
R_0T_0^2)}\left\{R_0(2T_0R'_0+R_0T'_0)-T_0(T_0R'_0+2R_0T'_0)\right.\right.\\\nonumber
&\times\left.\left.\frac{(1+2\lambda R_0^2T_0)}{(1+2\lambda
R_0T_0^2)}\right\}+\frac{\chi'_2}{(1+2\lambda
R_0T_0^2)}+\frac{2\lambda\chi_2}{(1+2\lambda
R_0T_0^2)}\left\{R_0(2T_0R'_0+R_0T'_0)\right.\right.\\\nonumber
&\times\left.\left.-\frac{2\lambda R_0^2T_0^2}{(1+2\lambda
R_0T_0^2)}\right\}+2\lambda
R_0(\chi_4-\chi_2)\frac{(2T_0R'_0+R_0T'_0)}{(1+2\lambda
R_0^2T_0)}+\frac{2\lambda R_0^2T_0\chi'_2}{(1+2\lambda
R_0^2T_0)}+(r\right.\\\nonumber &-\left.2m_0)\frac{(1+2\lambda
R_0^2T_0)}{(1+2\lambda
R_0T_0^2)}\left\{(\chi_4-\chi_5-\chi_2)\right\}\right]\eta=\frac{\omega^2}{(1+2\lambda
R_0^2T_0)}\left(\frac{b}{r^2}(1+2\lambda
R_0T_0^2)\right.\\\label{65}
&\times\left.(r+2m_0)(r-2m_0)+\overset{(P_2)}{\psi_{22}}\right)+\eta(\Phi+\Omega_{e})|_{pN}+(\chi_2+\chi_4)\frac{(r-2m_0)^2}{r^2}\varphi\eta.
\end{align}
For the onset of instability, one needs to satisfy above relation.
Since majority of the above terms are positive, therefore
instabilities will appear because of negative terms in the above
equation. For that reason we consider the following constraints to
be obeyed.
$$r>2m_0,~\chi_4>\chi_2+\chi_5,~2T_0R'_0+R_0T'_0)>T_0(T_0R'_0+2R_0T'_0)\frac{(1+2\lambda R_0^2T_0)}{(1+2\lambda
R_0T_0^2)}.$$ These are the required instability constraints that
the system must satisfy in order to enter in the unstable window
during evolution. In other words, the cylindrical relativistic
anisotropic interior will be unstable as long as it obeys above
relations. It is worthy to stress that in order to consider above
equation (\ref{65}) as an instable constraint, we need to consider
that all the terms on both sides of the equation are definite
positive. Thus, we take $\Phi>0$ and $\Omega_e>0$. It is seen that
this constraint depends merely on the static configurations of fluid
as well as $f(R,T)$ variables. Now, we consider N order effects,
then Eq.(\ref{65}) reduces to
\begin{align}\nonumber
&\frac{2\lambda\chi_4}{(1+2\lambda
R_0T_0^2)}\left\{R_0(2T_0R'_0+R_0T'_0)-\frac{2\lambda
R_0^2T_0^2}{(1+2\lambda
R_0T_0^2)}\right\}+r\left(1-\frac{2m_0}{r}\right)\\\nonumber
&\times\frac{(1+2\lambda R_0^2T_0)}{(1+2\lambda
R_0T_0^2)}(\chi_4-\chi_5-\chi_2)=(\chi_2+\chi_4)\left(1-\frac{2m_0}{r}\right)^2\varphi+
\left(1-\frac{2m_0}{r}\right)\\\label{66} &\times
\frac{\omega^2r}{(1+2\lambda R_0^2T_0)} \left\{{b}(1+2\lambda
R_0T_0^2)\left(1+\frac{m_0}{r}\right)+\overset{(P_2)}{\psi_{22}}\right\}+\Phi_N+\Omega_{e_N}.
\end{align}
Now, we use constant curvature condition firstly in above equation
and then in Eq.(\ref{24}). After using simultaneously these
equations, we get
\begin{align}\nonumber
&\left[1-\frac{2} {(1+2\lambda
\tilde{R_0}\tilde{T_0}^2)}\int^r_{r_{\Sigma^{(i)}}}\mu_0r^2dr+\frac{\lambda\tilde{R_0}^2\tilde{T_0}^2}
{r(1+2\lambda
\tilde{R_0}\tilde{T_0}^2)}\int^r_{r_{\Sigma^{(i)}}}r^2dr
\right]\frac{(1+2\lambda \tilde{R_0}^2\tilde{T_0})}{(1+2\lambda
\tilde{R_0}\tilde{T_0}^2)}\\\nonumber
&\times(\tilde{\chi}_4-\tilde{\chi}_5-\tilde{\chi}_2)
=\tilde{\varphi}\left[1-\frac{2} {(1+2\lambda
\tilde{R_0}\tilde{T_0}^2)}\int^r_{r_{\Sigma^{(i)}}}\mu_0r^2dr+\frac{\lambda\tilde{R_0}^2\tilde{T_0}^2}
{r(1+2\lambda \tilde{R_0}\tilde{T_0}^2)}\right.\\\nonumber
&\left.\times\int^r_{r_{\Sigma^{(i)}}}r^2dr
\right]^2(\tilde{\chi}_2+\tilde{\chi}_4)+\frac{\tilde{\omega}^2r}{(1+2\lambda
\tilde{R_0}^2\tilde{T_0})}\left[1-\frac{2} {(1+2\lambda
\tilde{R_0}\tilde{T_0}^2)}\int^r_{r_{\Sigma^{(i)}}}\mu_0r^2dr\right.\\\nonumber
&\left.+\frac{\lambda\tilde{R_0}^2\tilde{T_0}^2} {r(1+2\lambda
\tilde{R_0}\tilde{T_0}^2)}\int^r_{r_{\Sigma^{(i)}}}r^2dr \right]
\left[\overset{(P_2)}{\tilde{\psi}_{22}}+{b} \left\{1-\frac{2}
{(1+2\lambda
\tilde{R_0}\tilde{T_0}^2)}\int^r_{r_{\Sigma^{(i)}}}\mu_0r^2dr\right.\right.\\\label{67}
&-\left.\left.\frac{\lambda\tilde{R_0}^2\tilde{T_0}^2} {r(1+2\lambda
\tilde{R_0}\tilde{T_0}^2)}\int^r_{r_{\Sigma^{(i)}}}r^2dr
\right\}(1+2\lambda \tilde{R_0}\tilde{T_0}^2)
\right]+\frac{4\lambda^2\tilde{R_0}^2\tilde{T_0}^2\tilde{\chi_4}}{(1+2\lambda
\tilde{R_0}^2\tilde{T_0})^2}+\tilde{\Phi}_N+\tilde{\Omega}_{e_N},
\end{align}
where tilde shows that terms are evaluated under constant curvature
condition.  Now, we use ansatz $\mu_0={\xi}r^n$ in which
$n\in(-\infty, \infty)$, while $\xi$ is any positive real number
constant. This type of ansatz is well justified because any function
of a single variable can be expanded in a series form of that
variable. Considering $n\neq-3$ and $n\neq-4$, we get from the above
equation
\begin{align}\nonumber
&\left[1-\frac{2\xi(r^{n+3}-r^{n+3}_{{\Sigma^{(i)}}})}{3(1+2\lambda
\tilde{R_0}\tilde{T_0}^2)}
+\frac{2\xi(r^{n+4}-r^{n+4}_{{\Sigma^{(i)}}})}{3(n+4)(1+2\lambda
\tilde{R_0}\tilde{T_0}^2)}
+\frac{\lambda\tilde{R_0}^2\tilde{T_0}^2(r^{3}-r^{3}_{{\Sigma^{(i)}}})}{3r(1+2\lambda
\tilde{R_0}\tilde{T_0}^2)} \right](\tilde{\chi}_4\\\nonumber
&-\tilde{\chi}_5-\tilde{\chi}_2)\frac{(1+2\lambda
\tilde{R_0}^2\tilde{T_0})}{(1+2\lambda \tilde{R_0}\tilde{T_0}^2)}
=\left[1-\frac{2\xi(r^{n+3}-r^{n+3}_{{\Sigma^{(i)}}})}{3(1+2\lambda
\tilde{R_0}\tilde{T_0}^2)}+\frac{2\xi(r^{n+4}-r^{n+4}_{{\Sigma^{(i)}}})}{3(n+4)(1+2\lambda
\tilde{R_0}\tilde{T_0}^2)}\right.\\\nonumber
&+\left.\frac{\lambda\tilde{R_0}^2\tilde{T_0}^2(r^{3}-r^{3}_{{\Sigma^{(i)}}})}
{3r(1+2\lambda \tilde{R_0}\tilde{T_0}^2)}
\right]^2\tilde{\varphi}(\tilde{\chi}_2+\tilde{\chi}_4)+\frac{\tilde{\omega}^2r}{(1+2\lambda
\tilde{R_0}^2\tilde{T_0})}\left[1-\frac{2\xi(r^{n+3}-r^{n+3}_{{\Sigma^{(i)}}})}
{3(1+2\lambda \tilde{R_0}\tilde{T_0}^2)}\right.\\\nonumber
&+\left.\frac{2\xi(r^{n+4}-r^{n+4}_{{\Sigma^{(i)}}})}
{3(n+4)(1+2\lambda
\tilde{R_0}\tilde{T_0}^2)}+\frac{\lambda\tilde{R_0}^2\tilde{T_0}^2(r^{3}-r^{3}_{{\Sigma^{(i)}}})}
{3r(1+2\lambda \tilde{R_0}\tilde{T_0}^2)}\right] \left[{b}
\left\{1-\frac{2\xi(r^{n+3}-r^{n+3}_{{\Sigma^{(i)}}})}
{3r(1+2\lambda \tilde{R_0}\tilde{T_0}^2)}\right.\right.\\\nonumber
&+\left.\left.\frac{2\xi(r^{n+4}-r^{n+4}_{{\Sigma^{(i)}}})}
{3r(n+4)(1+2\lambda
\tilde{R_0}\tilde{T_0}^2)}-\frac{\lambda\tilde{R_0}^2\tilde{T_0}^2(r^{3}-r^{3}_{{\Sigma^{(i)}}})}
{r(1+2\lambda \tilde{R_0}\tilde{T_0}^2)} \right\}(1+2\lambda
\tilde{R_0}\tilde{T_0}^2)+\overset{(P_2)}{\tilde{\psi}_{22}}
\right]\\\label{68}
&+\frac{4\lambda^2\tilde{R_0}^2\tilde{T_0}^2\tilde{\chi_4}}{(1+2\lambda
\tilde{R_0}^2\tilde{T_0})^2}+\tilde{\Phi}_N+\tilde{\Omega}_{e_N}.
\end{align}
This is the required instability constraint at N era  for the
cylindrically symmetric collapsing systems framed within $f(R,T)$
gravity. This indicates that our relativistic instability constraint
depends on radial dependant fluid and $f(R,T)$ model variables. It
is worthy to stress that this instability constraint is independent
of stiffness parameter, which generally has utmost relevance in the
discussion dynamical instability of any stellar object. Here, we
also need to suppose that all quantities coming on the both sides of
the above equation are non-zero and non-negative.

\section{Summary and Discussion}

In this paper, we have explored some dynamical constraints which are
essential for a cylindrical object to be in a physically stable
state. We have constructed our analysis quiet systematically by
forming the modified field equations within the background of
$f(R,T)$ gravity theory. The cylindrical system is chosen to be
filled with anisotropic matter in the interior while the exterior
region is considered vacuum. The dynamical equations are obtained
from contracted Bianchi identities and some useful kinematical
variables are also explored including the expansion scalar. We have
presented the linear perturbation technique for metric and matter
variables with some known static profile of cylindrical object.
Initially, the system is assumed to be at rest and then gradually
enters into the non-static phase with the same time dependence on
the metric coefficients.

We have perturbed all the relevant equations to construct the
collapse equation. The general collapse equation is obtained by
using the conservation laws and field equations up to first order in
perturbation parameter. Also, we have constructed a real static
solution describing a collapsing state at large past time. For which
we have considered that all the metric functions in the static
background are positive indicating the cylindrical line element to
be Lorentz invariant. Moreover, we have categorized N, pN and ppN
eras by expanding the collapse equation up to order of
$\mathcal{C}^{-4}$.

For this purpose have converted our system to the c.g.s unit systems
because we know that every term related with some power of speed of
light have some physical interpretation and describes some useful
regimes. Since the terms associated with the zeroth order of speed
of light, i.e., $\mathcal{C}^0$ corresponds to N era and the terms
linked with $\mathcal{C}^{-2}$ provides the information of pN
regime. Similarly, the terms appearing with the order of
$\mathcal{C}^{-4}$ present the era of ppN. Further, we have imposed
the expansion free condition on the collapse equation due to
physical significance of this constraint. Such systems are
consistent with those astronomical objects which have an inner
cavity after the central explosion. A physical application of our
study is possible in those astrophysical objects which have cavity
in the interior region. This is due to the fact that cavity
formation in the the expansion free case is compensated by the
increment in the boundary surface during the overall expansion.

Generally, the adiabatic index ($\Gamma$) which describes the
rigidity in the fluid distribution indicate the instability regimes
for a gravitating source in the presence of expansion scalar.
Particular values of $\Gamma$ (i.e., $\Gamma<\frac{4}{3}$  for
spherical systems and $\Gamma<1$ for cylindrical objects with
perfect matter in the interior) exists in literature indicating the
unstable phase of the relativistic body. However, in the absence of
expansion scalar, this factor $\Gamma$ have no such importance to
evaluate the unstable regions of the relativistic system. To examine
the unstable regions of a self-gravitating cylindrical object with
zero expansion, it should satisfy the requirements (\ref{65}) and
(\ref{68}). The violation of these constraints describes the stable
configuration of the collapsing system. We found that the
instability range depends upon the dark source terms originating due
the $f(R,T)$ theory of gravity as well as on the length of the
cylinder. Moreover, the material profile like anisotropic pressure
and energy density also control the stability of the object during
the evolution.

The exploration of the zero expansion condition could be closely
linked with the study of voids which are sponge like structures and
can be explained with the Minkowskian cavity inside it. Thus, the
zero expansion condition asserts the existence of Minkowskian cavity
at the center. The potential applications of our dynamical analysis
are present in those astronomical objects which are carrying a
central Minkowskian cavity. Finally, we would like mention here that
all our results correspond to the instability constraints obtained
in GR under the particular limit, i.e., $f(R,T)=R$ \cite{13}.

\vspace{0.3cm}

\renewcommand{\theequation}{A\arabic{equation}}
\setcounter{equation}{0}
\section*{Appendix A}

The dark source terms $D_0$ and $D_1$ of Eqs.(\ref{18}) and
(\ref{19}) are given as follows
\begin{align}\nonumber
D_0&=\frac{\psi_{01,1}}{A^2}-\frac{\psi_{01}}{A^2}\left(\frac{A'}{A}+\frac{B'}{B}+\frac{2AA'}{B^2}
-\frac{CC'}{B^2}\right)-\frac{\psi_{00}}{A^2f_R}(B\dot{B}+C\dot{C})-\frac{\psi_{11}}{A^2f_R}\\\label{A1}
&\times B\dot{B}-\frac{\psi_{22}}{A^2f_R}C\dot{C},
\\\nonumber
D_1&=\frac{\psi_{01,0}}{B^2}-\frac{\psi_{01}}{B^2}\left(\frac{\dot{A}}{A}+\frac{\dot{B}}{B}
+\frac{2B\dot{B}}{A^2}
+\frac{C\dot{C}}{A^2}\right)-\frac{AA'}{B^2f_R}(\psi_{00}+\psi_{11})+\frac{CC'}{B^2f_R}\\\label{A2}
&\times(\psi_{11}-\psi_{22})-\left\{\frac{f-Rf_R}{2}-\psi_{11}\right\}_{,1}.
\end{align}
The components of extra curvature terms mentioned in
Eqs.(\ref{46})-(\ref{48}) are
\begin{align}\nonumber
\chi_2&=\frac{1}{B_0^2(1+2\lambda
R_0T_0^2)}\left\{\left(\frac{c}{r}\right)'\frac{B'_0}{B_0}
+\frac{1}{r}\left(\frac{b}{B_0}\right)'-\frac{c''}{r}\right\}-\left(\overset{(P)}{\psi_{00}}-e\lambda
R_0T_0^2\right)\\\nonumber &-\left(\mu_0-\lambda
R_0^2T_0^2+\overset{(S)}{\psi_{00}}\right)\left(\frac{2b}{B^2_0}+\frac{2\lambda
T_0^2}{(1+2\lambda R_0T_0^2)^3}\right),\\\nonumber
\chi_3&=\frac{1}{(1+2{\lambda}R_0^2T_0)}\left[(1+2{\lambda}R_0T_0^2)\left\{
\frac{A'_0}{A_0}\left(\frac{c}{r}\right)'\frac{1}{r}\left(\frac{a}{A_0}\right)\right\}
-\left\{\frac{\lambda}{2}{R_0^2}T_0^2+\overset{(S)}{\psi_{11}}\right.\right.\\\nonumber
&\left.\left.+(\mu_0+P_{r0})(1+2{\lambda}R_0^2T_0)
\right\}\left\{\frac{2b}{B_0^2} +2e\lambda
T_0^2\right\}-\left[2e\lambda R_0^2(\mu_0+P_{r0})+e\lambda
R_0T_0^2\right]\right],\\\nonumber
\chi_5&=-\frac{(1+2{\lambda}R_0T_0^2)}{A_0B_0^3(1+2{\lambda}R_0^2T_0)}
\left\{b'A'_0+a'B'_0-\frac{2b}{B_0}(A'_0B'_0)+B_0a''-bA''_0\right\}
-\left(2R_0P_{\phi0}\right.\\\nonumber
&+\left.2\mu_0R_0^2+T_0^2+\frac{\overset{(P)}{\psi_{22}}}{e\lambda
R_0}\right)\frac{e\lambda
R_0}{(1+2{\lambda}R_0^2T_0)}-(1+2{\lambda}R_0^2T_0)\left[
\frac{\lambda}{2}R_0^2T_0^2+\overset{(S)}{\psi_{22}}\right.\\\nonumber
&+\left.(\mu_0+P_{\phi0})
(1+2{\lambda}R_0^2T_0)\right]\left[\frac{b}{B_0}+\frac{a}{A_0}+\frac{2e\lambda
T_0^2}{(1+2{\lambda}R_0T_0^2)}\right].
\end{align}
The mathematical expressions mentioned in Eqs.(\ref{52}) and
(\ref{53}) are
\begin{align}\nonumber
\chi_1(r)&=\left[2\lambda T_0\mu_0\frac{(eT_0+2R_0z)}{(1+2\lambda
R_0T_0^2)}+\frac{(1+2\lambda R_0^2T_0)}{(1+2\lambda
R_0T_0^2)}(\mu_0+P_{r0})\left(\frac{c}{A_0^2}+\frac{bB_0}{A_0^2}\right)\right.\\\nonumber
&-\left.4\lambda\mu_0R_0\frac{(eT_0R_0z)}{(1+2\lambda
R_0^2T_0)}-\frac{z}{(1+2\lambda R_0^2T_0)}-\frac{1}{A_0^2(1+2\lambda
R_0T_0^2)}\left\{bB_0\overset{(S)}{\psi_{11}}\right.\right.\\\nonumber
&+\left.\left.
(bB_0+rc)\overset{(S)}{\psi_{00}}+rc\overset{(S)}{\psi_{22}}\right\}-\overset{(P_1)}{\psi_{01}}\frac{1}{A_0^2}
-\overset{(S)}{\psi_{01}}\frac{1}{A_0^2}\left(\frac{A'_0}{A_0}
+\frac{B'_0}{B_0}-\frac{r}{B_0^2}\right.\right.\\\label{A3}
&\left.\left.+\frac{2A_0A'_0}{B_0^2}\right)\right]\left[\frac{(1+2\lambda
R_0T_0^2)(1+2\lambda R_0^2T_0)}{(1+4\lambda
R_0^2T_0)}\right],\\\nonumber D_3&=\frac{2\lambda
R_0P_{r0}}{(1+2\lambda
R_0^2T_0)}(2T_0e'+2eR'_0+R_0z')-\frac{4e\lambda^2T_0P_{r0}}{(1+2\lambda
R_0^2T_0)^2}\left[P_{r0}(2T_0R'_0\right.\\\nonumber
&\left.+R_0T'_0)-T_0(T_0R'_0+2R_0T'_0)\frac{(1+2\lambda
R_0^2T_0)}{(1+2\lambda
R_0T_0^2)}\right]-\frac{4e\lambda^2T_0P_{r0}}{(1+2\lambda
R_0T_0^2)^2}(T_0R'_0\\\nonumber &+2R_0T'_0)\left\{\frac{(1+2\lambda
R_0^2T_0)}{(1+2\lambda R_0T_0^2)}-R_0^2\right\}+\frac{2\lambda
T_0P_{r0}}{(1+2\lambda
R_0T_0^2)^2}\left(T_0e'+2eR_0\frac{T'_0}{T_0}\right.\\\nonumber
&\left.+2R_0z'\right)(1+2\lambda
R_0^2T_0)-(\mu_0+P_{r0})\frac{A_0A'_0}{B_0^2(1+2\lambda
R_0T_0^2)}\left\{\frac{a}{A_0}+\frac{a'}{A'_0}-\frac{2b}{B_0}\right.\\\nonumber
&\left.-\frac{2e\lambda T_0}{(1+2\lambda
R_0T_0^2)}\right\}(1+2\lambda R_0^2T_0)-\frac{2e\lambda
T_0\mu'_0}{(1+2\lambda R_0T_0^2)^2}-\frac{2e\lambda
R_0^2A_0A'_0}{B_0^2(1+2\lambda R_0^2T_0)}\\\nonumber
&\times(\mu_0+P_{r0})+\frac{4e\lambda^2 T_0P_{r0}}{(1+2\lambda
R_0T_0^2)^2}\left[2\lambda
R_0^2T_0^2\frac{(T_0R'_0+2R_0T'_0)}{(1+2\lambda
R_0T_0^2)}-\mu_0(2T_0R'_0\right.\\\nonumber
&\left.+R_0T'_0)\right]+2\lambda
R_0\mu_0\frac{(2T_0e'+2eR'_0+R_0z')}{(1+2\lambda
R_0T_0^2)}+\frac{4\lambda^2T_0^2R_0^2\mu_0}{(1+2\lambda
R_0T_0^2)^2}\left(2R_0z'+T_0e'\right.\\\nonumber
&\left.+2eR_0\frac{T'_0}{T_0}\right)-4e\lambda^2T_0\mu_0R_0^2
\frac{(T_0R'_0+2R_0T'_0)}{(1+2\lambda
R_0T_0^2)^2}\left\{\frac{2\lambda T_0}{(1+2\lambda
R_0T_0^2)}-R_0^2\right\}\\\nonumber & + \frac{2e\lambda
R_0^2}{(1+2\lambda R_0^2T_0)}\left(\mu'_0+\frac{T'_0}{2}\right)
\left[1-\frac{2\lambda R_0^2T_0}{(1+2\lambda R_0^2T_0)}\right]
+\frac{2\lambda R_0^2T_0\bar{\mu}'}{(1+2\lambda
R_0^2T_0)}\\\nonumber &-\frac{\lambda R_0^2 T_0z'}{(1+2\lambda
R_0^2T_0)}+\frac{2\lambda R_0}{(1+2\lambda
R_0^2T_0)}(\mu_0-P_{r0})\left[\frac{2e\lambda R_0^2}{(1+2\lambda
R_0^2T_0)}(2T_0R'_0\right.\\\nonumber
&+\left.R_0T'_0)-2T_0e'-2eR'_0-R_0z'\right]+r\frac{(P_{r0}-P_{\phi0})}{B_0^2(1+2\lambda
R_0T_0^2)}(1+2\lambda R_0^2T_0)\left\{\frac{c}{r}\right.\\\nonumber
&\left.+c'-\frac{2e\lambda T_0}{(1+2\lambda
R_0T_0^2)}-\frac{2b}{B_0}\right\}-\frac{2er\lambda
R_0^2}{(1+2\lambda
R_0T_0^2)}(P_{r0}-P_{\phi0})-\left\{\left(\overset{(S)}{\psi_{00}}\right.\right.\\\nonumber
&\left.\left.+\overset{(S)}{\psi_{11}}\right)\left(\frac{a}{A_0}+\frac{a'}{A'_0}-\frac{2b}{B_0}
-\frac{2e\lambda T_0}{(1+2\lambda
R_0T_0^2)}\right)+\overset{(P)}{\psi_{00}}+\overset{(P)}{\psi_{11}}\right\}\frac{A_0A'_0}{B_0^2(1+2\lambda
R_0T_0^2)}\\\label{A4} &\lambda(eR_0
T-0^2)'+\overset{(P)}{\psi_{00,1}}+\frac{r}{B_0^2(1+2\lambda
R_0T_0^2)}\left(\overset{(P)}{\psi_{11}}-\overset{(P)}{\psi_{22}}\right).
\end{align}
The quantities $\Phi$ and $\Omega_e$ appearing in Eq.(\ref{64}) is
\begin{align}\nonumber
\Phi&=\frac{(r-2m_0)\omega^2}{(1+2\lambda
R_0T_0^2)}\left(\overset{(P_2)}{\psi_{11}}-\overset{(P_2)}{\psi_{22}}\right)
-\frac{\omega\varphi(r-2m_0)^2}{r^2(1+2\lambda
R_0T_0^2)}+\frac{\omega(r-2m_0)}{(1+2\lambda
R_0T_0^2)}\overset{(P_1)}{\psi_{11}}\\\nonumber &+\frac{2e\lambda
T_0 P'_{r0}}{(1+2\lambda R_0T_0^2)^2}+\frac{2e\lambda^2 T_0
P_{r0}}{(1+2\lambda
R_0T_0^2)^2}\left[P_{r0}(2T_0R'_0+R_0T'_0)-T_0(T_0R'_0\right.\\\nonumber
&\left.+2R_0T'_0)\frac{(1+2\lambda R_0^2T_0)}{(1+2\lambda
R_0T_0^2)}\right]+\frac{2\lambda R_0P_{r0}}{(1+2\lambda
R_0T_0^2)}(2T_0e'+2eR'_0+R_0z')-(T_0R'_0\\\nonumber
&+2R_0T'_0)\frac{4e\lambda^2T_0P_{r0}}{(1+2\lambda
R_0T_0^2)^2}\left\{\frac{(1+2\lambda R_0^2T_0)}{(1+2\lambda
R_0T_0^2)}-R_0^2\right\}+\frac{2\lambda T_0P_{r0}}{(1+2\lambda
R_0T_0^2)^2}\left(T_0e'\right.\\\nonumber
&\left.+2R_0z'+2eR_0\frac{T'_0}{T_0}\right)(1+2\lambda
R_0^2T_0)-\frac{(r-2m_0)^2}{r^2(1+2\lambda
R_0T_0^2)}(\mu_0+P_{r0})(1+2\lambda\\\nonumber &\times
R_0^2T_0)\left\{\frac{a}{r}(r+m_0)+\frac{m_0a'}{r^2}-\frac{2e\lambda
T_0}{(1+2\lambda R_0T_0^2)}\right\}-\frac{2e\lambda
R_0^2(r-2m_0)}{(1+2\lambda R_0^2T_0)}(\mu_0,\\\nonumber
&+P_{r0})-\frac{2e\lambda T_0\mu'_0}{(1+2\lambda
R_0T_0^2)}-\frac{4e\lambda^2T_0P_{r0}}{(1+2\lambda
R_0T_0^2)}[\mu_0(2T_0R'_0+R_0T'_0)]-2\lambda R_0^2T_0^2\\\nonumber
&\times\frac{(T_0R'_0+2R_0T'_0)}{(1+2\lambda
R_0T_0^2)}+\frac{2\lambda R_0\mu_0}{(1+2\lambda
R_0T_0^2)}(2T_0e'+2eR'_0+R_0z')+\frac{4\lambda^2
T_0^2R_0^2\mu_0}{(1+2\lambda R_0T_0^2)^2}\\\nonumber
&\times\left(T_0e'+2R_0z'+2eR_0\frac{T'_0}{T_0}\right)
-4e\lambda^2T_0\mu_0R_0^2\frac{(T_0R'_0+2R_0T'_0)}{(1+2\lambda
R_0T_0^2)^2}\left\{\frac{2\lambda T_0}{(1+2\lambda
R_0T_0^2)}\right.\\\nonumber &\left.-R_0^2\right\}+\frac{2e\lambda
R_0^2}{(1+2\lambda
R_0^2T_0)}\left(\mu_0'+\frac{T'_0}{2}\right)\left(1-\frac{2\lambda
R_0^2 T_0}{(1+2\lambda R_0^2T_0)}\right)-\frac{\lambda
T_0R_0^2z'}{(1+2\lambda R_0^2T_0)}\\\nonumber
&+\frac{2\chi'_2\lambda R_0^2T_0}{(1+2\lambda
R_0^2T_0)}-\frac{2\lambda R_0(\mu_0-P_{r0})}{(1+2\lambda
R_0^2T_0)}\left[2T_0e'+2eR'_0+2R_0z'-\frac{2e\lambda
R_0^2}{(1+2\lambda R_0^2T_0)}\right.\\\nonumber
&\left.\times(2T_0R'_0+R_0T'_0)\right]+\frac{(P_{r0}-P_{\phi0})}{(1+2\lambda
R_0T_0^2)}(r-2m_0)(1+2\lambda
R_0^2T_0)\left\{\frac{c}{r}+c'\right.\\\nonumber
&\left.-\frac{2e\lambda T_0}{(1+2\lambda
R_0T_0^2)}\right\}-\frac{2e\lambda R_0^2}{(1+2\lambda
R_0T_0^2)}(r-2m_0)(P_{r0}-P_{\phi0})-\frac{\varphi(r-2m_0)}{r^2(1+2\lambda
R_0T_0^2)}\\\nonumber
&\times\left\{\left(\overset{(S)}{\psi_{00}}+\overset{(S)}{\psi_{11}}\right)
\left(\frac{a}{r}(r+m_0)+\frac{m_0a'}{r^2}-\frac{2e\lambda
T_0}{(1+2\lambda
R_0T_0^2)}\right)+\overset{(P)}{\psi_{00}}+\overset{(P)}{\psi_{11}}\right\}\\\label{A5}
&\lambda(eR_0T_0^2)'+\overset{(P)}{\psi_{00,11}}+\frac{(r-2m_0)}{(1+2\lambda
R_0T_0^2)}\left(
\overset{(P)}{\psi_{00}}+\overset{(P)}{\psi_{22}}\right),\\\nonumber
\Omega_e&=\frac{2c}{r}\left[\frac{(P_{r0}-P_{\phi0})}{(1+2\lambda
R_0T_0^2)}(r-2m_0)(1+2\lambda
R_0^2T_0)\left\{\frac{c}{r}+c'-\frac{2e\lambda T_0}{(1+2\lambda
R_0T_0^2)}\right\}\right.\\\label{A6}
&\left.+\frac{\varphi(r-2m_0)}{(1+2\lambda
R_0T_0^2)}\left(\overset{(P)}{\psi_{11}}-\overset{(P)}{\psi_{22}}\right)
+\frac{\varphi(r-2m_0)}{r^2(1+2\lambda
R_0T_0^2)}(\mu_0+P_{r0})(1+2\lambda R_0^2T_0)\right].
\end{align}
The expansion of Eq.(\ref{g1}) provides
\begin{align}\nonumber
&P'_{r0}=\frac{1}{\mathcal{C}^0}\left[2\lambda
R_0^2T_0\frac{(P_{\phi0}-P_{r0})}{r}+2\lambda
P_{r0}T_0(T_0R'_0+2R_0T'_0) (1-2\lambda
R_0T_0^2)-256\right.\\\nonumber
&\left.\times\mathcal{G}m_0^2T_0m_0'-\left[\frac{2\lambda
m_0\mathcal{G}}{r^2l}(T_0R_0'+2R_0T'_0)\left\{2r^2l\mu_0-8
\lambda^2r^2lR_0^3T_0^3\mu_0-8\lambda^2r^3R_0^2T_0^2\right.\right.\right.\\\nonumber
&\times\left.\left.\left.\mu_0l-32\lambda
r^2P_{r0}R_0T_0^2-32r^3P_{r0}\lambda
m_0T_0\right\}\right]_{,1}-4\lambda^2\left[\frac{R_0T_0^2}{r^2l}(T_0R'_0+2R_0T'_0)\left\{4\lambda
r^2\right.\right.\right.\\\nonumber
&\times\left.\left.\left.lR_0^2T_0\mu_0
\mathcal{G}+16r^2P_{r0}m_0\mathcal{G}^2-64\lambda^2R_0^3T_0^3P_{r0}\mathcal{G}^2r^2
-64\lambda^2r^3R_0^2T_0^2P_{r0}\mathcal{G}^2\right\}\right]_{,1}
+{\lambda^2 T_0^3}\right.\\\nonumber
&\times\left.\left(2R_0T'_0+T_0R''_0+2R_0T''_0+2R_0\frac{T'^2_0}{T_0}\right)4rR_0P_{r0}\mathcal{G}-\frac{\lambda
T_0m_0}{rl\mathcal{G}}-\frac{4m_0\lambda
T_0}{rl}\left(2R_0T'_0\right.\right.\\\nonumber
&+\left.\left.T_0R''_0+2R_0T''_0+2R_0\frac{T'^2_0}{T_0}\right)\left(4r^2\lambda
lR_0^2T_0\mu_0\mathcal{G}+16r^2m_0\mathcal{G}^2P_{r0}-64
\lambda^2R_0^3T_0^3P_{r0}r^2\mathcal{G}^2\right.\right.\\\nonumber
&-\left.\left.64\lambda^2r^3(\mathcal{G}R_0T_0)^2P_{r0}\right)
+\frac{4m_0^2T_0\lambda}{r^2l}(16\mathcal{G}r^2\mu_0m_0
-64\lambda^2R_0^3T_0^3r^2\mu_0\mathcal{G}-64m_0r^3(T_0\right.\\\nonumber
&\times\left.R_0\lambda)^2\mu_0\mathcal{G}+2r^2lP_{r0}-8\lambda^2R_0^3T_0^4r^2lP_{r0}
-8\lambda^2r^3T_0^2R_0^2lP_{r0})+\frac{\mathcal{G}\mu_0}{2rl}(2r^2l\mu_0\right.\\\nonumber
&-\left.8\lambda^2r^2lR_0^3T_0^3\mu_0-8\lambda^2r^3R_0^2T_0^2\mu_0l-32\lambda
r^2P_{r0}R_0T_0^2-32r^3P_{r0}\lambda
m_0T_0)+\frac{\mathcal{G}}{2rl}\right.\\\nonumber
&\times\left.\left(\frac{P_{r0}}{\mathcal{G}}+2\lambda
R_0^2T_0\mu_0+2\lambda
R_0^2T_0P_{r0}\mathcal{G}-\frac{4m_0\mu_0}{r}\right)\left(\frac{\lambda
R_0^2T_0^2r^2l}{\mathcal{G}}-32r^2\mu_0\lambda
R_0T_0^2\right.\right.\\\nonumber &-\left.\left.32\mu_0\lambda
m_0rT_0-\frac{4\lambda
lP_{r0}R_0T_0^2}{\mathcal{G}}-\frac{4r^3\lambda l
P_{r0}T_0}{\mathcal{G}}+8\lambda
(R_0T_0r)^2m_0\right)\right]\\\nonumber
&+\frac{\mathcal{G}}{\mathcal{C}^2}\left[\frac{2m_0}{r}(P_{r0}-P_{\phi0})-8\lambda^3
P_{r0}T_0^4(T_0R'_0+2R_0T'_0)+2\lambda
\mu_0R_0(2T_0R'_0+R_0T'_0)\right.\\\nonumber &+\left.4\lambda
\mu_0R_0^2T_0^2(R'_0T_0+2R_0T'_0)-256m_0^2\mathcal{G}m_0'-\left[\frac{2\lambda
m_0}{r^2l}(T_0R'_0+2R_0T'_0)\{16r^2\mathcal{G}\mu_0m_0\right.\right.\\\nonumber
&-\left.\left.64\lambda^2R_0^3T_0^3r^2\mu_0\mathcal{G}-64m_0r^3T_0^2\lambda^2R_0^2\mu_0\mathcal{G}
+2r^2lP_{r0}-8\lambda^2R_0^3T_0^3P_{r0}r^2T_0l-8r^3\right.\right.\\\nonumber
&\times\left.\left.\lambda^2T_0^2R_0^2lP_{r0}\}\right.]'
-16\lambda^3R_0^2T_0^4(T_0R'_0+2R_0T'_0)P_{r0}\mathcal{G}+4\lambda^2R_0P_{r0}
T_0^3m_0\left(2R_0T'_0\right.\right.\\\nonumber
&+\left.\left.2R_0T''_0+T_0R''_0+2R_0\frac{T'^2_0}{T_0}\right)+\frac{4m_0^2T_0\lambda}{r^2l
\mathcal{G}}(4r^2lR_0^2T_0\mu_0\mathcal{G}+16r^2P_{r0}m_0P_{r0}\mathcal{G}^2\right.\\\nonumber
&-\left.64\lambda^2R_0^3T_0^3r^2\mathcal{G}^2P_{r0}-64\lambda^2r^3R_0^2T_0^2P_{r0}\mathcal{G}^2)
+\frac{\mu_0}{2rl}\{16r^2\mathcal{G}\mu_0m_0-64\lambda^2R_0^3T_0^3r^2\mathcal{G}\mu_0\right.\\\nonumber
&-\left.64m_0r^3T_0^2\lambda^2R_0^2\mu_0\mathcal{G}+2r^2lP_{r0}-8\lambda^2R_0^3T_0^3r^2lT_0P_{r0}
-8\lambda^2r^3T_0^2R_0^2lP_{r0}\}+\frac{1}{2r^2l}\right.\\\nonumber
&\times\left.\left(\frac{P_{r0}}{\mathcal{G}}+2\lambda
R_0^2T_0\mu_0+2\lambda
R_0^2T_0P_{r0}\mathcal{G}-\frac{4m_0\mu_0}{r}\right)(2r^2l
\mu_0-8\lambda^2r^2lR_0^3T_0^3\mu_0-8\lambda^2\right.\\\nonumber
&\times\left.r^3R_0^2\mu_0T_0^2l-32\lambda
r^2P_{r0}R-0T_0^2-32r^3\lambda
P_{r0}m_0T_0)+\frac{1}{r^2l}(-4\lambda
R_0^2T_0m_0-2m_0\right.\\\nonumber
&\times\left.P_{r0}-4m_0R_0^2T_0\lambda
P_{r0}\mathcal{G})(-32r^2\mu_0\lambda R_0T_0^2-32\mu_0\lambda
m_0rT_0-4\lambda l P_{r0}R_0T_0^2\mathcal{G}^{-1}\right.\\\nonumber
&-\left.4r^3\lambda l P_{r0}T_0\mathcal{G}^{-1}+\lambda
R_0^2T_0^2r^2l\mathcal{G}^{-1}+8\lambda
R_0^2T_0^2r^2m_0)\right]\\\nonumber
&+\frac{\mathcal{G}}{\mathcal{C}^4}\left[(P_{\phi0}-P_{r0})\mathcal{G}^{-1}-2\lambda
P_{r0}R_0(2T_0R'_0+R_0T'_0)+2\lambda
P_{r0}T_0(T_0R'_0+2R_0T'_0)\right.\\\nonumber
&-\left.4\lambda^2P_{r0}T_0^2(T_0R'_0+2R_0T'_0)R^2_0+4\lambda^2\mu_0^2R_0\mathcal{G}(2T_0R'_0
+R_0T'_0)-4\lambda^2R_0^2T_0\mu'_0\mathcal{G}\right.\\\nonumber
&-\left.\left\{\frac{2\lambda m_0}{r^2l}(T_0R'_0+2R_0T'_0)(4\lambda
r^2lR_0^2T_0\mu_0\mathcal{G}+16r^2lP_{r0}\mathcal{G}^2m_0l^{-1}-64\lambda^2R_0^3T_0^3r^2\right.\right.\\\nonumber
&\times\left.\left.P_{r0}\mathcal{G}^2-64\lambda^2r^3R_0^2T_0^2\mathcal{G}^2P_{r0})\right\}'
+16m_0^2T_0^3\lambda^2\mathcal{G}R_0P_{r0}+\frac{1}{2rl}(P_{r0}\mathcal{G}^{-1}+2\lambda
R_0^2\right.\\\nonumber &\times\left.T_0\mu_0+2\lambda
R_0^2T_0P_{r0}\mathcal{G}-4m_0\mu_0r^{-1})(16\mathcal{G}r^2\mu_0m_0-64\lambda^2(R_0T_0)^3r^2\mu_0\mathcal{G}
-64\right.\\\nonumber &\times\left.m_0r^3T_0^2\lambda^2R_0^2\mu_0
\mathcal{G}+2r^2lP_{r0}-8\lambda^2(R_0T_0)^3r^2lT_0P_{r0}
-8\lambda^2r^3T_0^2lT_0^2P_{r0})+\frac{1}{r^2l}\right.\\\nonumber
&\times\left. (-4\lambda R_0^2T_0m_0-2m_0P_{r0}-4m_0R_0^2+\lambda
T_0P_{r0}\mathcal{G})(2r^2l\mu_0-8\lambda^2r^2l(R_0T_0)^3\mu_0\right.\\\nonumber
&-\left.8\lambda^2r^3(T_0R_0)^2\mu_0l-32\lambda
r^2P_{r0}R_0T_0^2-32r^3P_{r0}\lambda m_0T_0)\right].
\end{align}

\section*{Acknowledgments}

We would like to thank  Professor Malcolm A. H. MacCallum for his
many valuable suggestions and comments that significantly improved
the paper. This work was supported by Higher Education Commission,
Islamabad through start up research grant program.

\end{document}